# Amorphous and Nanocrystalline Topological Semimetal YPtBi/W/CoFeB Heterostructures for BEOL-Compatible Spin-Orbit Torque Devices

Quang Le[1], Brian R. York[1], Cherngye Hwang[1], Xiaoyong Liu[1], Tsann Lin[1], Xiaoyu Xu[1], Yudi Wang[1], Jia Li[1], Mazin Osman[1], Katherine Le[1], Maher Osman[1], Son Le[1], Lei Xu[1], Maki Maeda[2], Tuo Fan[2], Yu Tao[2], Hisashi Takano[2], Sho Kagami[3], Ohiro Fujie[3], and Pham Nam Hai[3]

[1] Western Digital Corp., Great Oaks, San Jose, California 95119, USA

[2] Western Digital Corp., Fujisawa, Kanagawa 252-0811, Japan

[3] Department of Electrical and Electronic Engineering, Institute of Science Tokyo, Meguro, Tokyo 152-8552, Japan

## ABSTRACT

Spin-orbit torque (SOT) devices require spin-source materials that combine efficient charge-to-spin conversion with back-end-of-line (BEOL) thermal compatibility. Here, we show that YPtBi/W/CoFeB heterostructures deposited directly on Si/$SiO_x$ remain predominantly amorphous or weakly nanocrystalline from room temperature to 400 °C while preserving a large effective damping-like SOT response. Anomalous Hall and harmonic Hall measurements, together with X-ray diffraction, cross-sectional transmission electron microscopy, X-ray reflectivity, and electron energy-loss spectroscopy, show that the response does not correlate with bulk crystallization of YPtBi. Instead, the interfacial analysis indicates that the strongest trend of the spin Hall angle is associated with the chemistry of the upper YPtBi/W boundary: the effective SOT response tracks the integrated W concentration at that YPtBi surface. Meanwhile, a two-spin source analysis shows that the Pt-W-rich interlayer provides only a small positive correction, insufficient to explain the large negative effective spin Hall angle by itself. The dominant control variable is therefore inferred to be the incorporation of W into the upper YPtBi interface, which plausibly modifies the local electronic structure of YPtBi and amplifies the stack-level response. These results provide a more physically constrained interpretation of the stack behavior and identify a BEOL-compatible route to disordered topological spin-source layers for scaled SOT memory and compute-in-memory hardware.

## I. INTRODUCTION

Spin-orbit torque magnetic random-access memory (SOT-MRAM) is attractive for embedded and stand-alone memory applications because its three-terminal geometry separates read and write current paths, thereby reducing stress on the magnetic tunnel junction barrier and enabling fast switching. For technology insertion, however, a useful SOT stack must deliver more than a large nominal spin torque efficiency. It must also provide that response at practical resistivity, within a realistic CMOS-compatible thermal budget, and with sufficient microstructural uniformity to limit device-to-device write-current spread across arrays. Recent reviews, therefore, frame SOT insertion as a coupled materials, transport, and interface problem rather than as a search for a single maximum spin Hall angle [1-3].

The half-Heusler topological semimetal YPtBi is an especially interesting candidate because sputtered films have shown large effective spin Hall responses at the wide deposition temperature range of 300-600°C [4-7], while the broader material family also exhibits nontrivial semimetallic and topological surface-state physics [8-10]. At the same time, experiments on topological insulators such as BiSb have implicitly linked high SOT efficiency to crystallinity [11], texture control [12], or growth optimization on single-crystalline substrates [13,14]. That assumption is problematic for manufacturing. If a strong response depends on a narrow crystallographic process window, the material is inherently more difficult to integrate on large 300 mm wafers. In addition, aggressive grain growth and texture development can introduce local variations in resistivity, current partitioning, and surface morphology. From a device-integration perspective, a disordered topological material without long-range order is therefore not automatically inferior. The relevant question is whether local coordination and the correct interfacial electronic structure can preserve strong charge-to-spin conversion without requiring bulk long-range order.

This view is reinforced by recent momentum-resolved studies of amorphous and nanocrystalline topological materials have shown that short-range order can preserve coherent, momentum-dependent topological electronic states even after long-range translational symmetry is lost [15]. Such works provide an important physical precedent: disorder does not necessarily eliminate electronically and spintronically relevant topological surface states.

The central question addressed in this study is therefore whether a technologically useful SOT response can be maintained in YPtBi topological semimetal-based heterostructures without relying on bulk crystallization. In this work, we focus on YPtBi/W/CoFeB-based stacks deposited on $Si/SiO_x$ and evaluate their structural and transport response over a deposition temperature range from room temperature to 450 °C. The working hypothesis is that the decisive variable is not the onset of long-range order throughout the full YPtBi layer, but the evolution of the upper YPtBi interface into a chemically mixed and electronically active region.

The spin Hall angle is conventionally written as

$$\theta_{\mathrm{SH}} = \left(\frac{2e}{\hbar}\right)\left(\frac{J_{\mathrm{s}}}{J_{\mathrm{c}}}\right) \tag{1}$$

where $J_{\mathrm{s}}$ is the perpendicular spin-current density and $J_{\mathrm{c}}$ is the in-plane charge-current density in the SOT layer. In the present study, the experimentally relevant quantity is the damping-like SOT efficiency $\xi_{\mathrm{DL}}$ (or the effective spin Hall angle $\theta_{\mathrm{SH}}^{\mathrm{eff}}$) extracted from harmonic Hall measurements. Note that the harmonic Hall analysis does not measure a purely intrinsic value of the SOT layer's spin Hall angle. The extracted damping-like efficiency is an effective value that depends on intrinsic spin generation, spin transmission across interfaces, and the magnetic boundary conditions of the entire stack [16]. Following the standard low-frequency harmonic Hall formalism [17], the effective damping-like field $H_{\mathrm{DL}}$ is related to the drive current through

$$\theta_{\mathrm{SH}}^{\mathrm{eff}} = \left(\frac{2e}{\hbar}\right)\left(\frac{\mu_0 M_{\mathrm{s}} t_{\mathrm{eff}} H_{\mathrm{DL}}}{J_{\mathrm{c}}}\right), \tag{2}$$

where $M_{\mathrm{s}}$ is the saturation magnetization of the ferromagnet, $t_{\mathrm{eff}}$ is the effective ferromagnet thickness, and $J_{\mathrm{c}}$ is the current density assigned to the SOT layer. As mentioned above, $\theta_{\mathrm{SH}}^{\mathrm{eff}}$ extracted in this way is a stack-level effective quantity. It depends on intrinsic spin-current generation in the SOT layer, transmission of spin current across the YPtBi/W/CoFeB interfaces, and the electronic structure of the active interfacial region. For that reason, structural, compositional, and transport measurements must be interpreted together if the origin of the measured response is to be physically constrained.

That distinction matters for integration. If a strong SOT response requires bulk crystallization of the YPtBi topological semimetal, the deposition temperature itself becomes the dominant design constraint. If, instead, the decisive variable is the chemistry and density of the upper YPtBi interface, then a lower-deposition-temperature engineering route becomes available for the full SOT stacks.

## II. EXPERIMENTAL METHODS

Multilayers with the nominal structure substrate/buffer/YPtBi (100 Angstrom) / W (8 Angstrom) / CoFeB (12 Angstrom)/ MgO-based cap were deposited on Si/$SiO_2$ substrates by physical vapor deposition. The YPtBi spin-source layer was formed by co-sputtering as in Ref. 4. However, we kept the substrate temperature much lower than in a previous work [18]. The W interlayer was inserted to suppress intermixing and diffusion between YPtBi and CoFeB. The representative experimental matrix and the corresponding characterization flow are summarized in Table 1. Here, the cold-cap and hot-cap splits correspond to the thermal condition used during formation of the oxide buffer/cap. The purpose of this split was to determine whether the spin-torque response is governed primarily by the stack's overall thermal history or persists across different oxide-processing environments, provided that the active interfacial region remains disordered.

*Table 1. Representative experimental matrix used to organize the process splits and characterization flow discussed in the Results and Discussion.*

| Split | YPtBi deposition window | Primary purpose | Key methods |
|---|---|---|---|
| Cold-cap | RT-400 °C | Amorphous-window torque retention | AHE, harmonic Hall, XRD, TEM, XRR |
| Cold-cap | 450 °C | Crystallization onset and post-amorphous comparison | AHE, XRD, TEM |
| Hot-cap | RT-400 °C | Process sensitivity of torque retention under alternate oxide condition | AHE, harmonic Hall, XRD, XRR |
| Hot-cap | 450 °C | Crystallized-state benchmark | AHE, XRD, TEM |

Patterned Hall structures were fabricated for anomalous Hall effect (AHE) and second harmonic Hall measurements. Structural evolution was examined by X-ray diffraction (XRD) and cross-sectional transmission electron microscopy (TEM). X-ray reflectivity (XRR) was used to estimate thickness, roughness, and fitted layer density, because these quantities help distinguish simple topographic broadening from the formation of a physically denser mixed region. Electron energy-loss spectroscopy (EELS) was used to assess the local composition near the active interface. The spin Hall effect was evaluated by the second harmonic Hall technique. Sheet-resistance mapping was further used to estimate current partitioning across the multilayer stack, enabling the correct extraction of the spin torque efficiency by Eq. (2).

Figure 1 summarizes the stack architecture, the patterned Hall geometry, a representative anomalous Hall loop, and the corresponding second harmonic Hall signal used throughout the study.

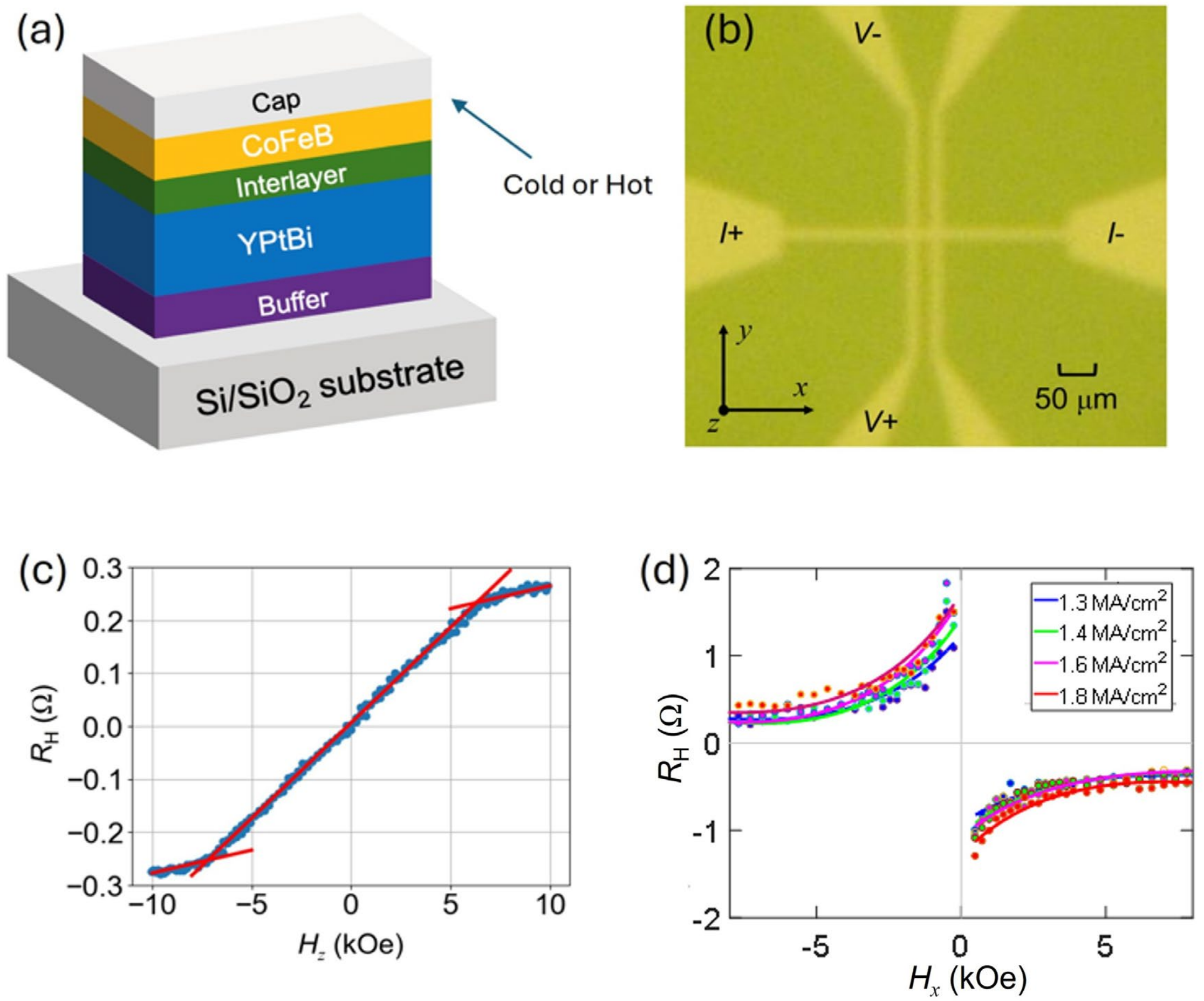

*Figure 1. Device and measurement overview. (a) Schematic multilayer stack on $Si/SiO_2$, (b) optical image of the patterned Hall device, (c) representative anomalous Hall loop measured with a perpendicular magnetic field, and (d) representative harmonic Hall response measured with an in-plane magnetic field used to extract the effective SOT signal.*

## III. RESULTS AND DISCUSSION

### 3.1 Disordered structural window of YPtBi from room temperature to 400 °C

The first question is whether the YPtBi layer remains structurally disordered over the same thermal window in which the stack preserves useful magnetic transport. Figure 2 addresses this point by combining the out-of-plane and in-plane XRD datasets for different deposition temperatures. No strong YPtBi diffraction peak is resolved in samples deposited at temperatures from room temperature to 400 °C, implying an amorphous or extremely fine nanocrystalline state. Although XRD alone cannot distinguish a fully amorphous film from nanocrystallites that remain below its detection limit, it does show that the long-range order required to produce well-developed YPtBi reflections is absent throughout the studied thermal window.

The 450 °C deposition temperature serves as an important structural boundary condition. Once the deposition temperature enters this regime, the diffraction response changes qualitatively, and texture signatures emerge, indicating the onset of a more crystallized state. Furthermore, we find that the texture in the crystallized state depends on the cap deposition temperature. The hot-cap samples exhibit a different (220) orientation tendency from the cold-cap samples, which show a mixture of (111) and (220) orientations. These results argue against a unique crystallographic orientation preference as the origin of the strong effective torque observed below 400 °C, as discussed later. Figure 2 indicates that YPtBi thin films deposited at low temperatures below 400 °C lack strong diffraction peaks. Nevertheless, we observe a large spin Hall effect in this structurally disordered regime.

TEM and FFT images provide the complementary local-structure evidence needed to sharpen that interpretation. As shown in Figure 3, the room-temperature and 350 °C samples display predominantly diffuse FFT contrast, consistent with short-range order (SRO), and lack the extended lattice-fringe continuity expected of a highly crystallized YPtBi layer with long-range order (LRO). The 400 °C sample remains predominantly disordered but contains a limited population of larger grains embedded within a broader amorphous or weakly nanocrystalline matrix. In other words, 400 °C represents the upper edge of a disordered window rather than the onset of a fully crystallized YPtBi phase.

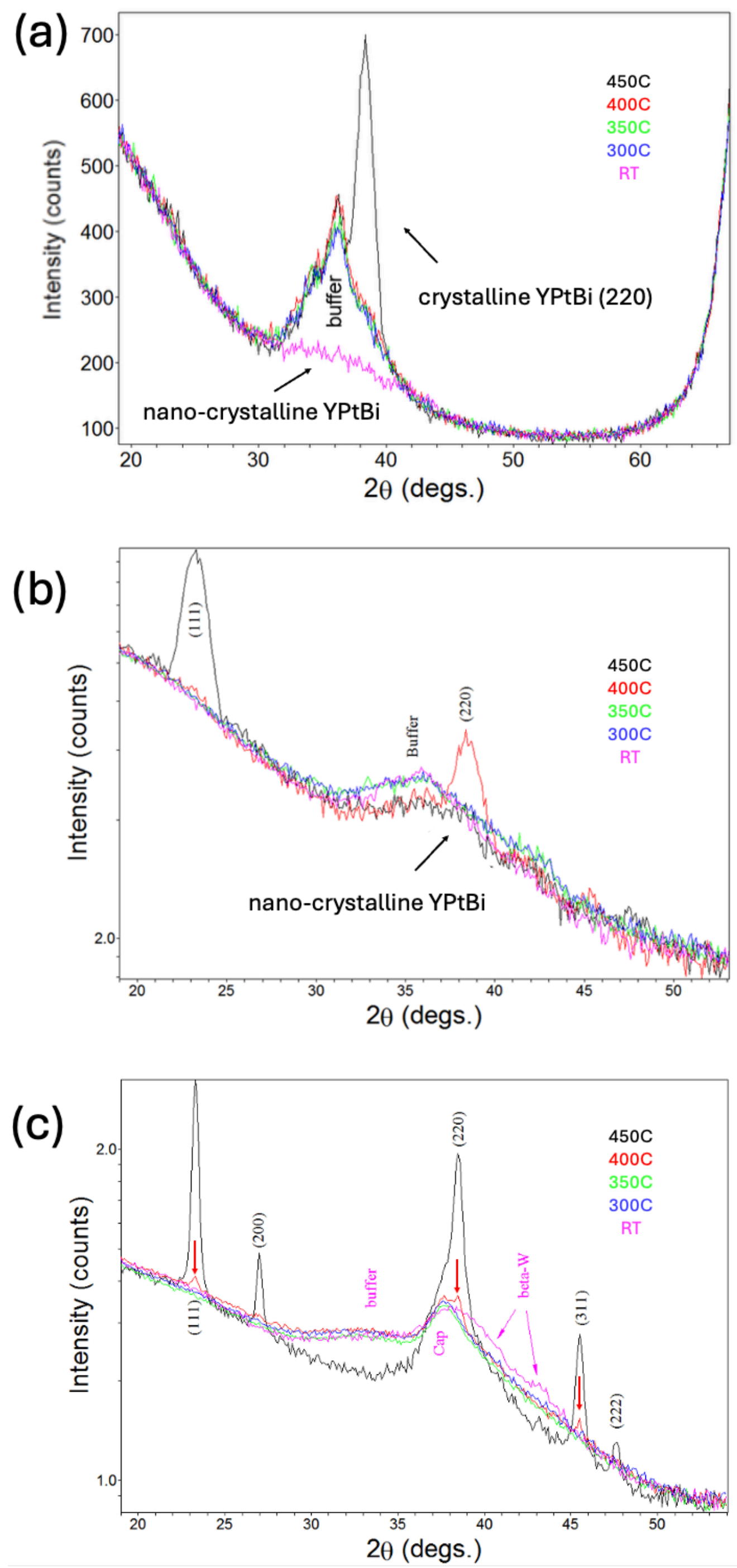

***Figure 2***. *XRD evidence for the disordered thermal window. (a) Out-of-plane XRD for hot-cap samples, (b) out-of-plane XRD for the cold-cap samples, and (c) in-plane XRD showing that only limited crystalline contributions (red arrows) appear near 400 °C deposition temperature, whereas the 450 °C deposition temperature exhibits a much stronger crystalline signature.*

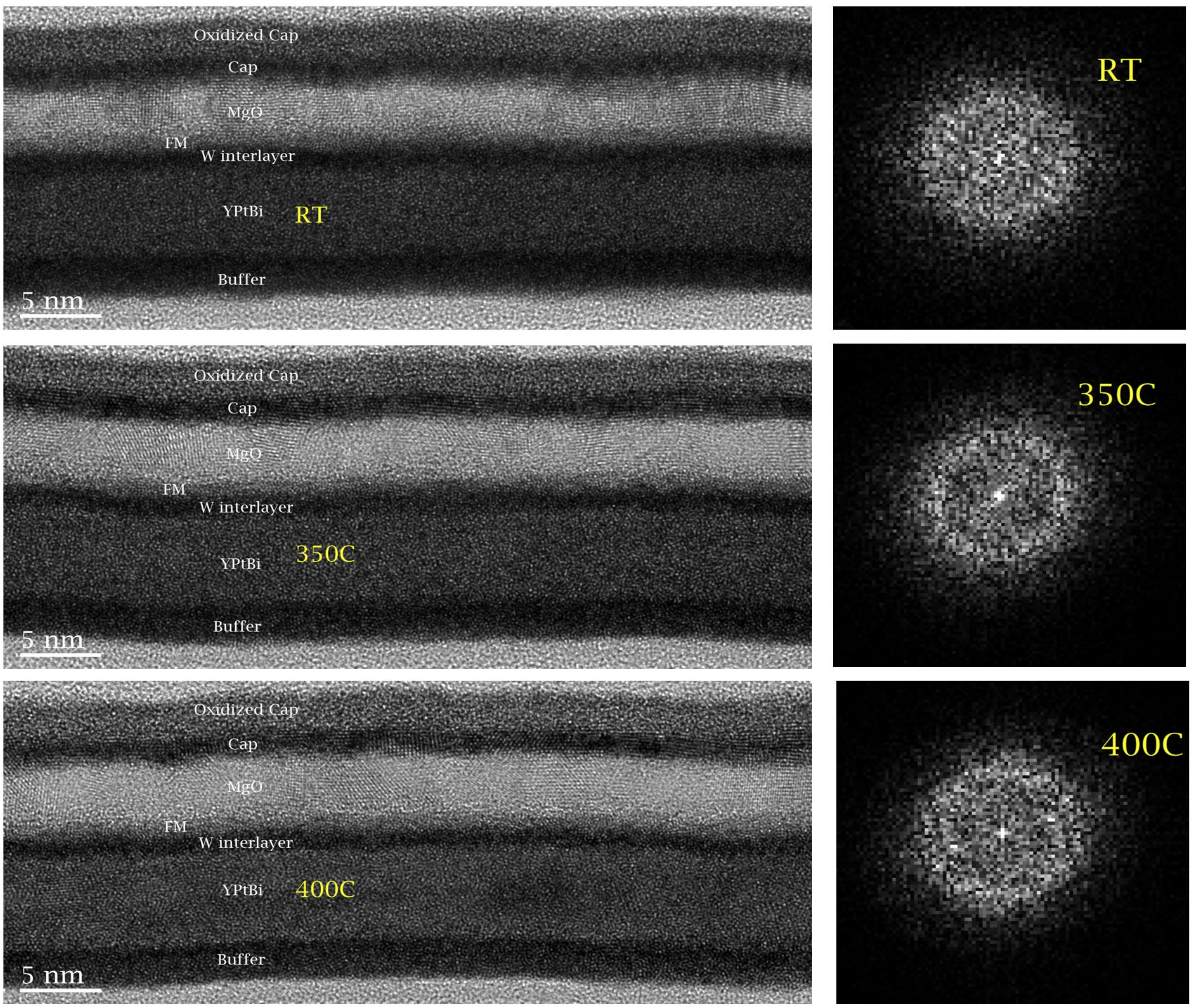

***Figure 3.*** *Cross-sectional TEM and corresponding FFT evidence for local disorder. Representative datasets show that YPtBi is predominantly amorphous from room temperature to 400 °C. At 400 °C, isolated weakly nanocrystalline regions appear locally, but the layer remains dominated by a broader amorphous or weakly nanocrystalline matrix.*

Read together, Figs. 2 and 3 establish the structural regime for interpreting the subsequent transport data. From room temperature to 400 °C, YPtBi is best described as a mixed amorphous or weakly nanocrystalline phase, with short-range order preserved, consistent with the broader idea that local order can remain electronically relevant even in the absence of bulk translational periodicity [9]. That conclusion is technically important because it implies that any mechanism that requires bulk, long-range crystallinity to generate a large spin response is inconsistent with the measured temperature dependence.

## 3.2 Magnetotransport response and persistence of the amorphous-window spin-orbit torque

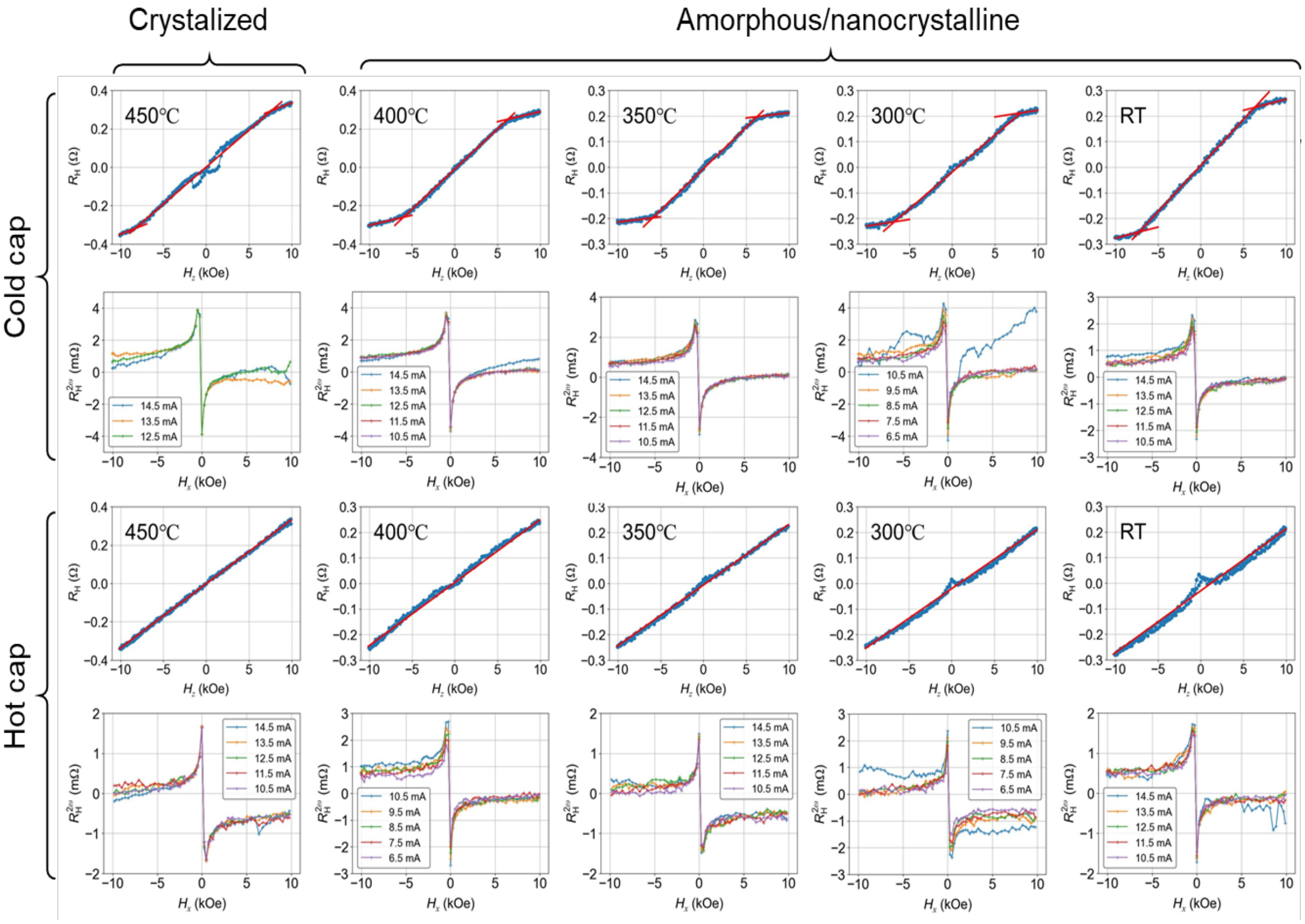

***Figure 4****. Magnetotransport of YPtBi/W/CoFeB heterostructures. Upper panels show anomalous Hall loops; lower panels show second harmonic Hall data used to extract the effective damping-like response. A large effective SOT signal persists throughout the structurally disordered thermal window.*

With the structural window established, the next issue is whether the stack retains a strong effective SOT response within the same temperature range. Figure 4 compiles AHE loops and harmonic Hall signals for the cold-cap and hot-cap process families. The AHE loops remain well-defined from room temperature to 400 °C, indicating that the ferromagnetic response is preserved while the YPtBi layer remains structurally disordered. The corresponding second-harmonic signals show a negative spin Hall angle under the present sign convention and remain sizable in both process families. This contrasts previous works which show that YPtBi has a positive spin Hall angle.

The important point is not merely that a response is measurable. Rather, the large stack-level response persists before the onset of full crystallization. In the hot-cap branch, for example, the magnitude is already large at the low-temperature end of the process window and remains strong through 400 °C. This behavior argues against a simple picture in which the spin-orbit torque appears only after bulk YPtBi ordering develops. Instead, the disordered state already contains the electronically active configuration needed for efficient charge-to-spin conversion.

### 3.3 Process-family trends across electrical, magnetic, and torque-related metrics

Figure 5 summarizes the sheet-resistance trend, the saturation magnetization $M_{\mathrm{s}}$, the magnetic anisotropy field $H_{\mathrm{k}}$, the effective spin Hall angle (SHA), and the anomalous Hall resistance as a function of YPtBi deposition temperature for the cold-cap and hot-cap process families. These data provide a process-summary view of how transport, magnetic state, and extracted torque-related metrics co-evolve across the two deposition conditions.

Several trends are technically informative. First, the hot-cap family exhibits a generally lower sheet resistance response across much of the temperature range explored, indicating a more conductive stack. Second, the same branch retains larger $M_{\mathrm{s}}$ and $H_{\mathrm{k}}$ over a broader portion of the process window, indicating that the magnetic layer and adjacent interfaces remain more magnetically robust. Third, the anomalous Hall amplitude is broadly consistent with that picture, particularly at the higher deposition temperatures where the hot-cap branch preserves a larger signal.

The spin-orbit torque metric is more nuanced. The effective spin Hall angle does not scale monotonically with resistance alone, which argues against a purely conductivity-driven $\theta_{SH}{\sim}1/R$. Lower resistance can coexist with better magnetic retention, but it does not by itself determine the final effective SHA. This distinction becomes central in the following structural and chemical analysis, which reveal that the decisive variable is the local interfacial state of the YPtBi/W/CoFeB boundary region.

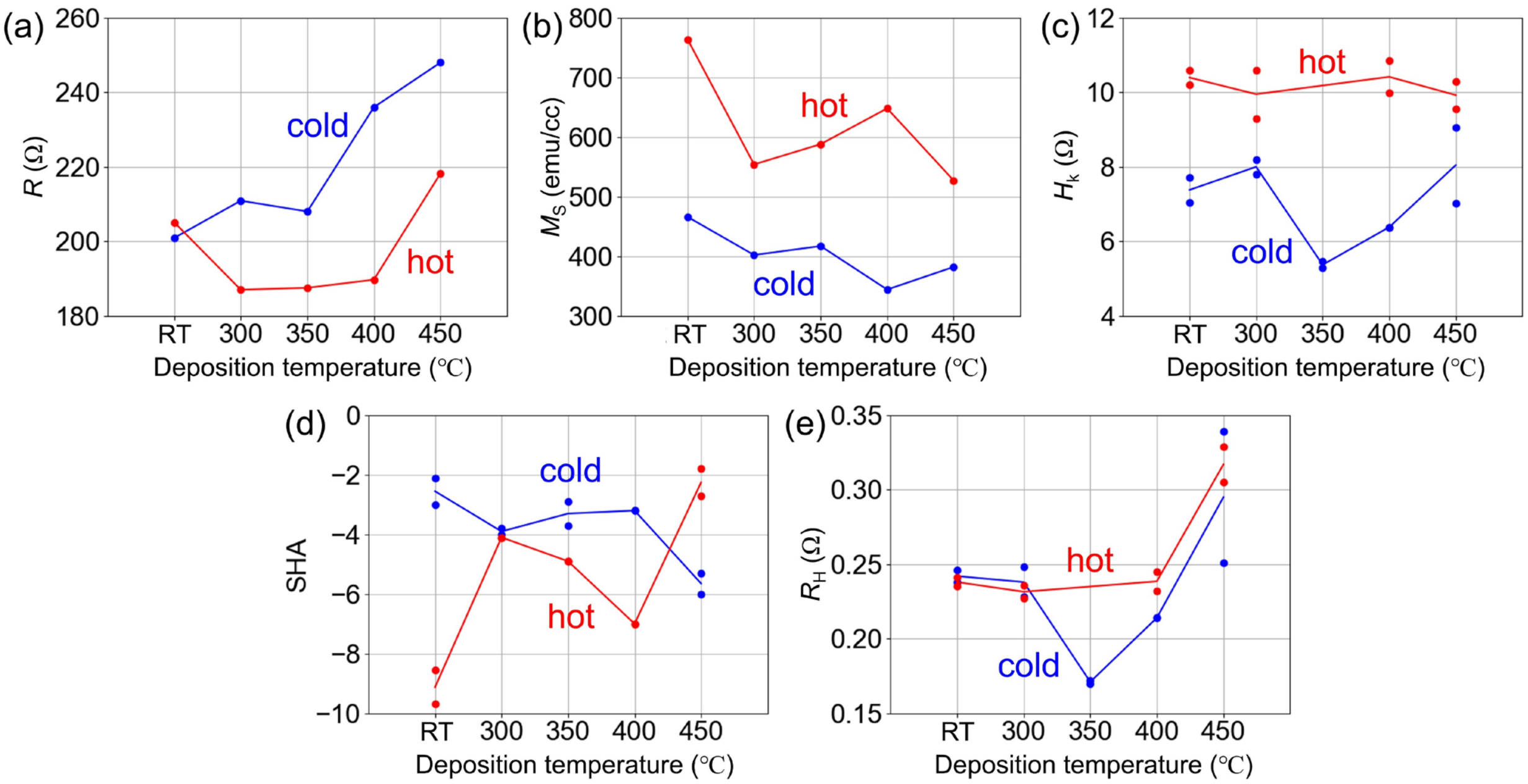

***Figure 5.*** *Process-family summary trends versus YPtBi deposition temperature: (a) resistance, (b) saturation magnetization, (c) magnetic anisotropy field, (d) effective spin Hall angle, and (d) anomalous Hall resistance amplitude. The hot-cap branch generally preserves stronger magnetic properties and lower resistance, whereas the effective torque metric shows a more selective dependence on the interfacial state.*

### 3.4 Interface-localized structural evolution from cross-sectional TEM and XRR

Having established that bulk YPtBi crystallization is not required for a strong spin-orbit torque response, we now determine where in the stack the relevant evolution occurs. Figure 6 shows cross-sectional TEM images of the active interface region. The YPtBi layer remains largely disordered through the relevant temperature window, but the boundary among YPtBi, the nominal

W interlayer, and the Co-containing ferromagnetic region is visibly broadened. The broadened zone is localized near the interface, suggesting that the governing process is interfacial mixing rather than homogeneous modification of the entire YPtBi layer.

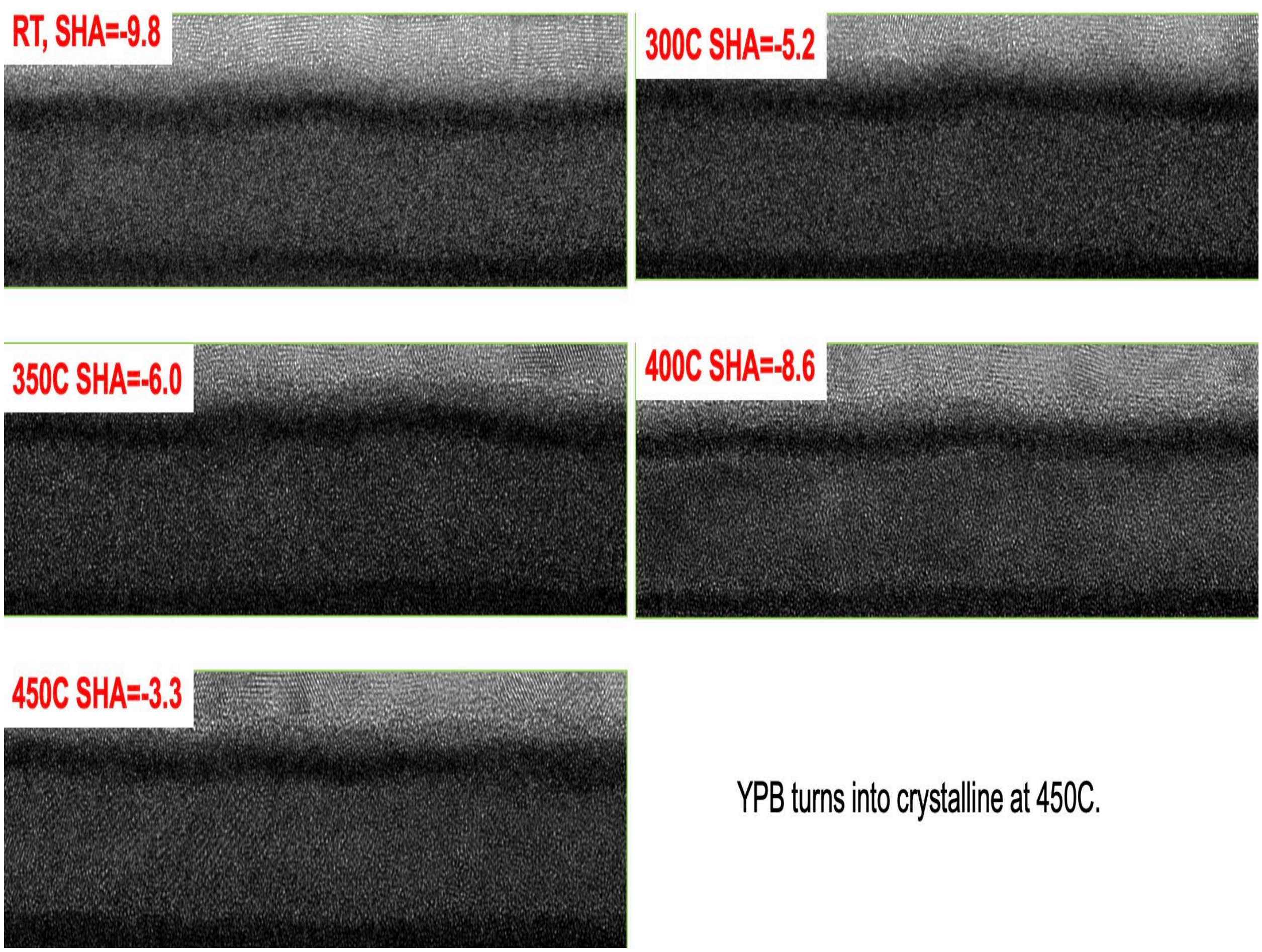

***Figure 6.*** *Cross-sectional structural mapping of the active interface. The YPtBi region remains largely disordered through the relevant process window, while the YPtBi/W/Co-containing boundary exhibits measurable interfacial broadening and mixing.*

X-ray reflectivity was then used to estimate the evolution of the average layer density and topographic roughness. Figure 7(a) summarizes the fitted layer roughness values across the multilayer stack. The largest roughness variation is not located at the CoFeB/W/YPtBi region itself, but rather at the MgO/CoFeB interface. By contrast, Figure 7(b) shows that the fitted density of the nominal W-derived region above YPtBi tracks the magnitude of the negative effective response much more strongly.

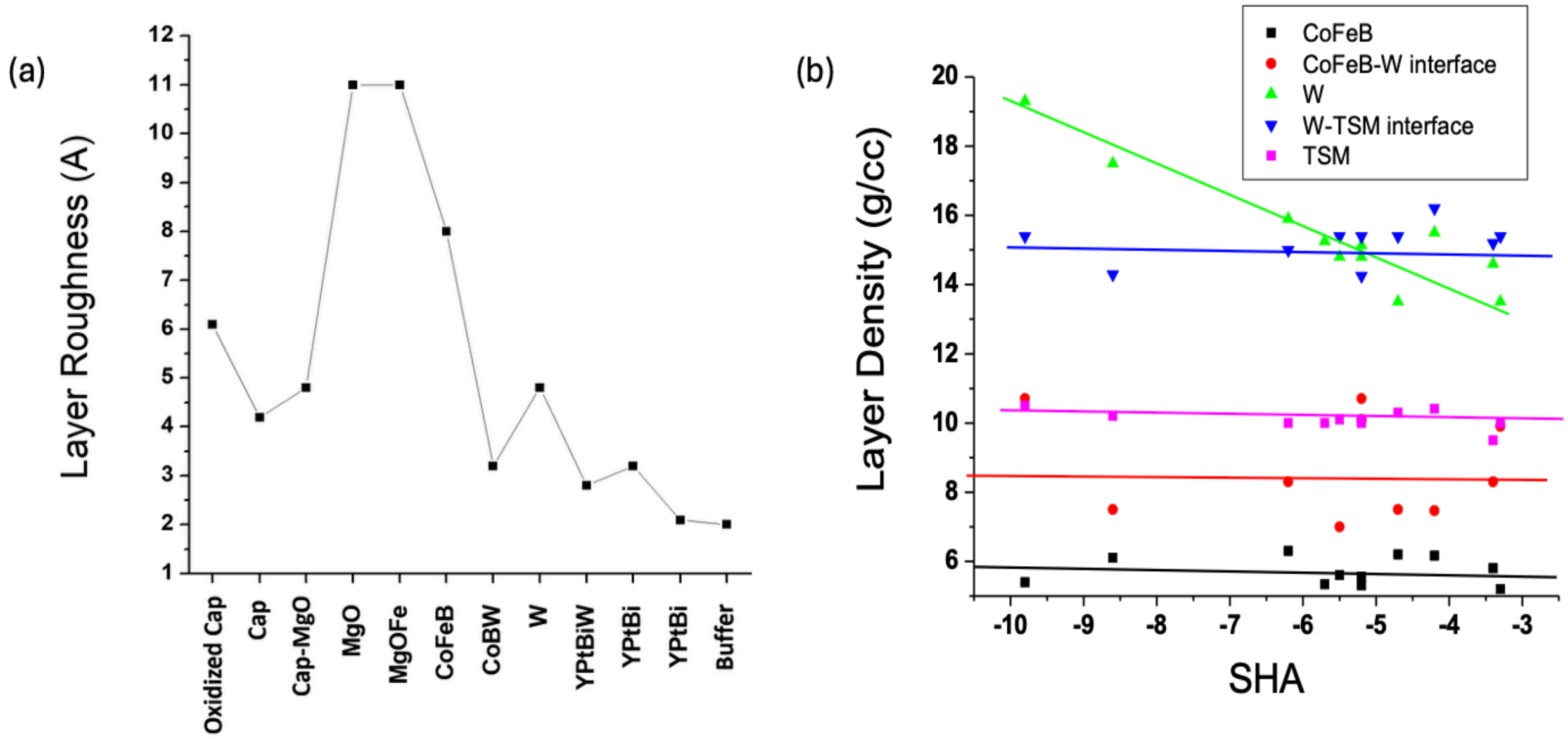

*Figure 7. XRR-derived structural trends. (a) Fitted layer roughness across the stack and (b) fitted layer-density trends showing that the nominal W-derived region above YPtBi tracks the measured effective response more strongly than bulk YPtBi roughness does.*

Density in this context should not be overinterpreted as a unique phase identifier. The fitted density value can reflect intermixing, alloying, compaction, or a chemically modified interfacial region. The important point is narrower and more physical: the variable that best tracks the measured response is a localized structural-property change at the W/YPtBi interface, rather than texture development within the 100-Angstrom-thick bulk YPtBi. This result already points to an interface-controlled mechanism before any detailed compositional analysis is invoked.

### 3.5 Interfacial composition, Pt and W redistribution, and a physically consistent mechanism

To further probe the vertical chemical asymmetry of the nominal W-derived interlayer, a four-segment EELS analysis was performed across its thickness as shown in Fig. 8. This analysis clearly showed that the lower part of the W interlayer, adjacent to YPtBi, is relatively Pt-rich, whereas the upper part, adjacent to CoFeB, is relatively Co-rich. The result is therefore consistent with a chemically graded mixed interlayer rather than a compositionally uniform interlayer. Figure 8 also shows that the YPtBi surface region is compositionally graded with measurable W penetration into the upper YPtBi surface. The boundary between YPtBi/W has therefore evolved into a mixed YPtBi(W) / W(Pt) region rather than remaining an abrupt, compositionally pure sequence of layers.

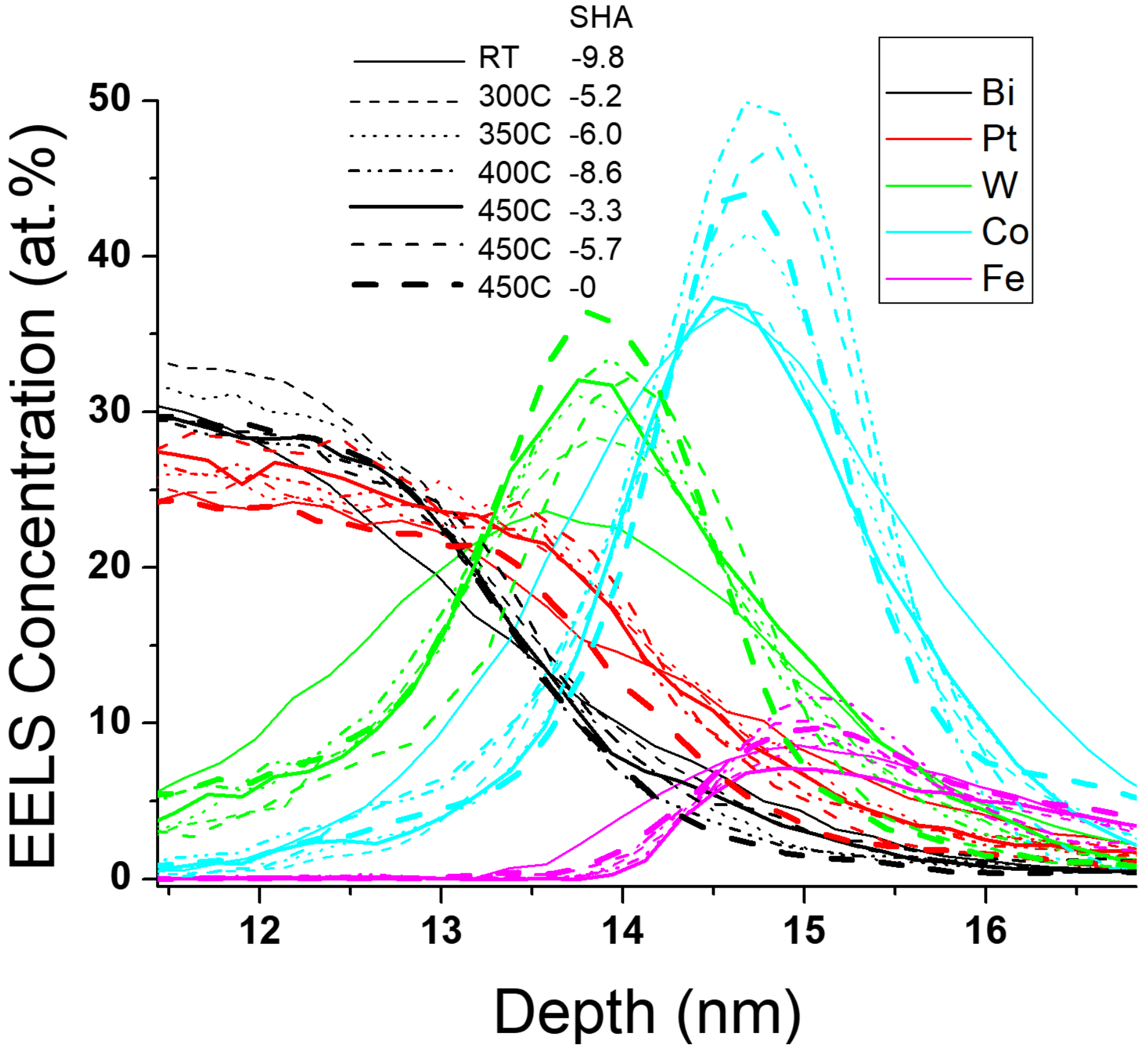

***Figure 8.*** *EELS depth profiles of the active YPtBi/W/CoFeB interfacial region. The profiles indicate a graded mixed layer with Pt redistribution upward from YPtBi and Co redistribution downward from the ferromagnetic side within the nominal W-derived interlayer. The Bi profile marks the top of the TSM region.*

This observation provides a physically consistent interpretation of the earlier structural and transport results. EELS depth profiles show Pt diffusion into the nominal W interlayer. Because Pt and W are much denser than Y and Bi, a higher fitted density in the nominal W-derived interlayer is naturally consistent with Pt enrichment or Pt-W alloying in that region.  We therefore compared the EELS-derived Pt concentration relative to W, averaged across the full width at half maximum of the nominal interlayer, to the XRR-fitted W-layer density. Figure 9 summarizes the correlation identified in the interfacial chemistry analysis. A higher average Pt fraction, corresponding to a more Pt-rich interlayer, exhibits higher fitted density, consistent with the intrinsic density of Pt and with the formation of a structurally modified Pt-W-rich interlayer.

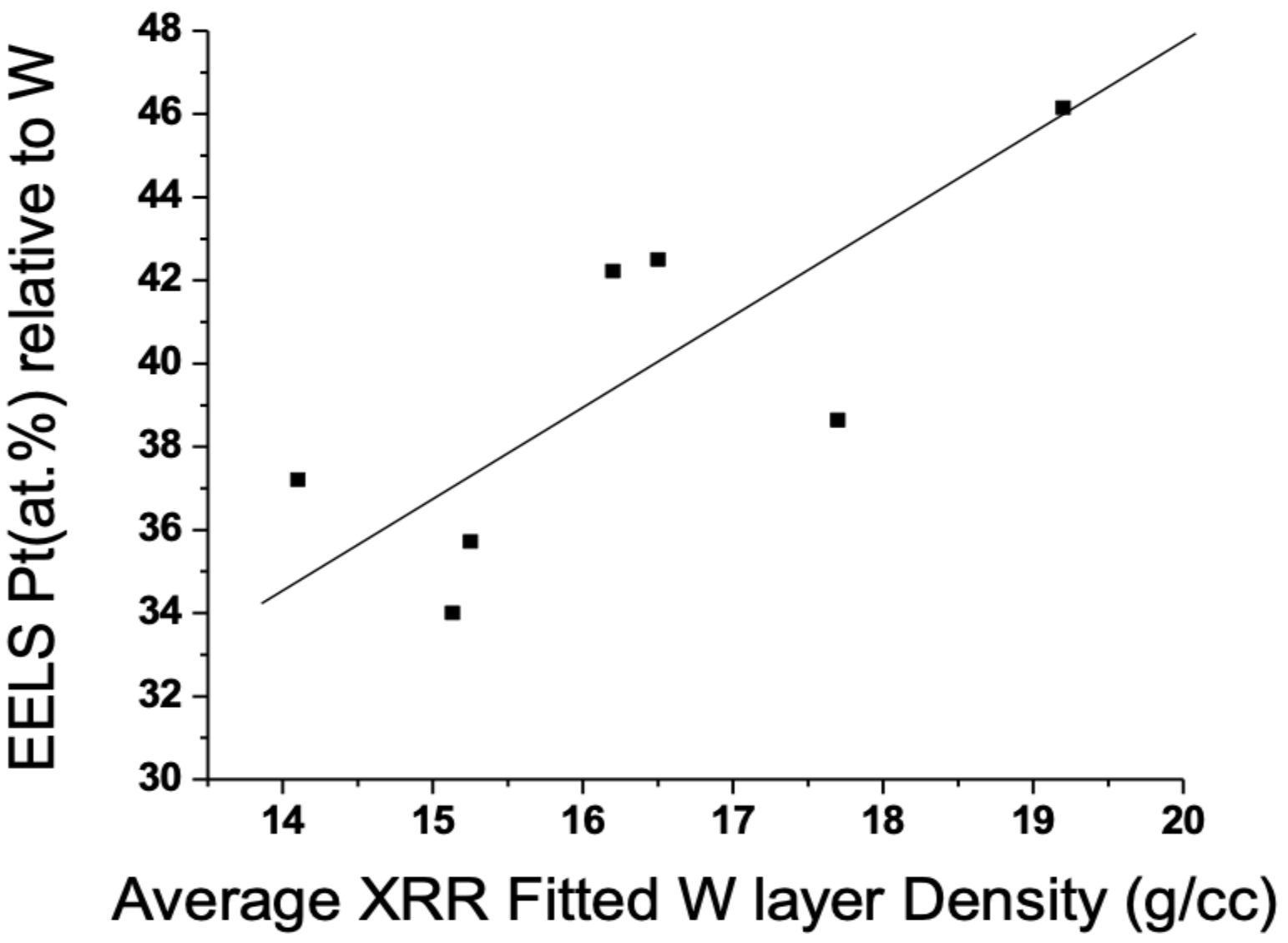

*Figure 9. Relative Pt content to W in the W(Pt) interlayer versus fitted W-interlayer density.*

### 3.6 Origin of the effective spin Hall angle

The correlation in Fig. 7(b) and the chemical trends in Figs. 8 and 9 suggest that the interlayer W(Pt) may contribute to the observed effective spin-orbit torque. We first examine this possibility using a two-source framework adapted from Eq. (5) of Kagami et al. for YPtBi/Ta/CoFeB heterostructures [7]. For the present YPtBi/W/CoFeB system, the intrinsic contribution is associated with the spin-charge conversion that originates from YPtBi, whereas the parasitic contribution is associated with the Pt-containing nominal W(Pt) interlayer,

$$\theta_{\mathrm{SH}}^{\mathrm{eff}} \approx \theta_{\mathrm{SH}}^{\mathrm{YPtBi}} \operatorname{sech}\left(\frac{t_{\mathrm{W(Pt)}}}{\lambda_{\mathrm{W(Pt)}}}\right) + \left(\frac{\sigma_{\mathrm{W(Pt)}}}{\sigma_{\mathrm{YPtBi}}}\right) \theta_{\mathrm{SH}}^{\mathrm{W(Pt)}}. \quad (3)$$

Here, $\theta_{SH}^{YPtBi}$ denotes the YPtBi-derived spin source term that reaches the magnetic interface, $t_{W(Pt)}$ and $\lambda_{W(Pt)}$ are the effective thickness and spin-diffusion length of the mixed interlayer, and $\left(\frac{\sigma_{W(Pt)}}{\sigma_{YPtBi}}\right)\theta_{SH}^{W(Pt)}$ represents the interlayer contribution weighted by current partitioning. In the conventional sign convention, Pt exhibits a positive spin Hall angle, whereas W exhibits a negative one [19,20]. However, analysis of the present composition range using spin-Peltier-based Pt–W alloy data and alloy-composition studies [21,22] indicates that the interlayer compositions relevant here, approximately 50–65% W and 35–50% Pt, fall in a regime where the effective spin Hall angle of the Pt–W alloy remains predominantly positive, thus the parasitic spin Hall effect from W(Pt) cannot account for the large negative $\theta_{SH}^{eff}$ observed in our stacks. When thickness dependence of the spin Hall angle is also taken into account, the effective SHA of an approximately 8 Å Pt-containing W interlayer is small and positive on the order of 0.005–0.015, while the corresponding conductivity is approximately $5\times10^5$–$8\times10^5$ S/m. With $\sigma_{YPtBi}$ ~ $1\times10^5$ S/m, the quantity $\left(\frac{\sigma_{W(Pt)}}{\sigma_{YPtBi}}\right)\theta_{SH}^{W(Pt)}$ is therefore positive but modest in magnitude (~ 1% of the magnitude of the observed $\theta_{SH}^{eff}$).

This result has an important consequence for the interpretation of the stack-level spin-orbit torque data. Since a small positive Pt–W-related term cannot explain either the large negative $\theta_{SH}^{eff}$ or its variation across the RT–400 °C process window, the dominant control variable must therefore reside elsewhere. Taken together with the XRR and EELS results, the most plausible explanation is the intrinsic contribution of the chemically evolved YPtBi(W) surface region itself. The W incorporation into the upper YPtBi interface can modify the local electronic structure of the topological surface states, which then sets the sign and magnitude of the measured spin-orbit torque response. Because Dirac-dispersion topological surface states and Berry-curvature-driven spin-charge conversion in YPtBi are expected to be sensitive to the local band filling and the position of the Fermi level relative to the Dirac manifold, one plausible interpretation is that W incorporation and Pt depletion due to interdiffusion on the upper YPtBi interface shifts the interfacial electronic structure toward a more favorable condition for spin-current generation with increasing W concentration and decreasing Pt concentration in the YPtBi(W) surface. In this revised interpretation, the mixed Pt–W interlayer remains relevant as a transport- and spin-transparency-modifier, but the large and negative $\theta_{\mathrm{SH}}^{\mathrm{eff}}$ is governed primarily by the electronically modified YPtBi(W) interface.

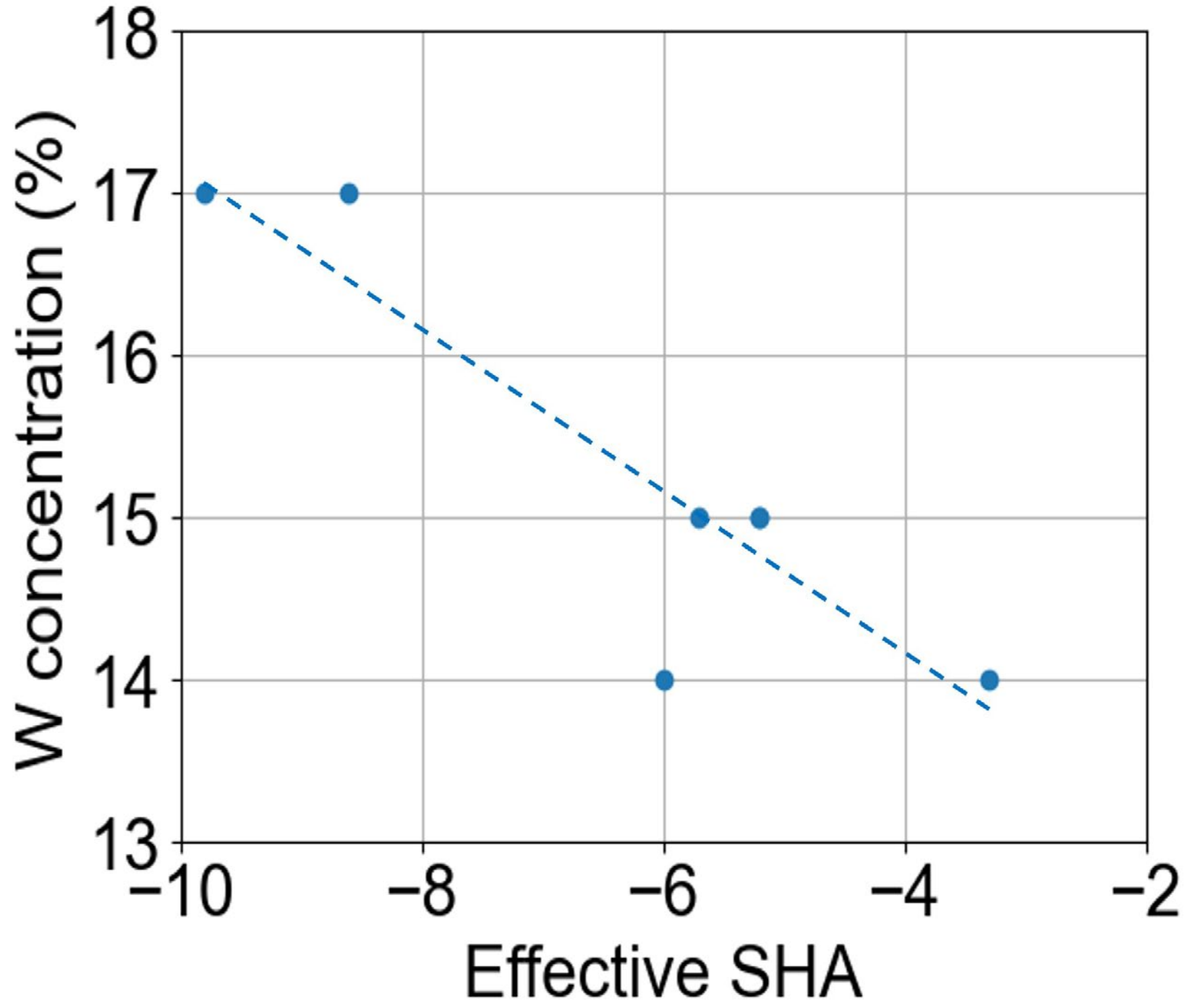

***Figure 10***. *W concentration integrated over the upper YPtBi(W) surface versus effective SHA. The monotonic trend indicates that higher interfacial W concentration is associated with a more negative effective response, supporting the interpretation that W incorporation into the upper YPtBi interface is more directly connected to the stack-level spin-orbit torque.*

Figure 10 provides an additional test of this hypothesis by plotting the W concentration integrated over the YPtBi(W) surface as a function of $\theta_{\mathrm{SH}}^{\mathrm{eff}}$. Although the data show some scatter, the trend is monotonic: samples with higher integrated interfacial W concentration tend to exhibit a more negative effective response. This behavior supports the view that chemistry within the upper YPtBi(W) interface is more directly linked to the sign and magnitude of the measured spin-orbit torque. The correlation is therefore consistent with, though not by itself definitive proof of, an interface-doping mechanism in which W incorporation at the upper YPtBi surface modifies the local electronic structure and amplifies the Dirac-derived topological surface states and Berry-curvature-driven spin-charge conversion in YPtBi.

### 3.7 Implications for device integration

Previous works on high spin Hall angles of YPtBi requires that the YPtBi layer was deposited at a high temperature of 300~600 °C on sapphire substrates for good crystallization [4-7]. Furthermore, amorphous non-topological YPt thin films do not show large spin Hall angles [23]. From an integration perspective, a disordered topological YPtBi spin-source layer can be advantageous if it preserves spin-orbit torque efficiency while reducing the need for epitaxy or highly specific texture control. That point is especially relevant for BEOL processing, where the practical thermal ceiling is typically near 400 °C and where process latitude often matters as much as peak material performance [2]. The present results indicate that a useful design strategy for YPtBi-based SOT devices is to engineer the top YPtBi interface and the adjacent interlayer chemistry rather than to force bulk crystallization across the full thickness of the topological semimetal.

That interpretation is important because localized boundary evolution can strongly alter a measured stack-level spin-orbit torque without requiring bulk reordering. Changes in interfacial charge-to-spin conversion efficiency by the chemistry of the topological surface, interfacial spin transparency, and spin-flip scattering occur at length scales much smaller than the full thickness of the YPtBi layer, and the consequent spin-orbit torque can therefore change substantially even if the broader YPtBi region remains amorphous or only weakly nanocrystalline. This interpretation is also favorable from a variability standpoint. If the dominant control variable is a local YPtBi interface within an otherwise disordered film, it may be possible to reduce grain-structure-driven variability relative to a strongly textured or coarse-grained design. While the present work does not yet prove array-level manufacturability; that will require wafer-level statistics on device-to-device switching current, magnetic tunnel-junction performance, and endurance, our results do identify a practical materials direction for integration of this half-Heusler topological material to SOT devices.

## IV. CONCLUSION

A YPtBi/W-based SOT heterostructure deposited directly on Si/SiOx maintains a large negative effective spin Hall response while remaining predominantly amorphous or weakly nanocrystalline through the 400 °C BEOL window. The expanded structural analysis shows that the high-response regime precedes full crystallization of YPtBi and therefore does not require a fully ordered YPtBi film. Instead, the strongest updated correlations identify the upper YPtBi/W interface as the critical region: the effective response tracks the integrated W concentration at that boundary more directly than either of the bulk YPtBi structural orderings.

Taken together, the XRD, TEM, XRR, EELS, and magnetotransport results support a physically constrained picture in which disordered YPtBi provides the baseline spin-source functionality, while W incorporation into the upper YPtBi interface acts as the dominant control variable for the large negative effective response. Meanwhile, two-source analysis shows that the W(Pt) interlayer contributes a positive but comparatively small correction and is therefore insufficient to explain the measured trend on its own. A plausible interpretation is that W incorporation and Pt depletion modify the local electronic structure of YPtBi near the top interface, potentially changes the relevant Dirac-derived states relative to the Fermi level and thereby enhancing the magnitude and polarity of the effective response. Although that electronic-structure hypothesis remains to be tested directly via first principle calculations using the density-functional theory, the present data establish interface engineering at the top of disordered YPtBi, rather than enforced bulk crystallization, as the more relevant materials route for BEOL-compatible YPtBi-based SOT devices.

**Conflict of Interest:** The authors have no conflicts to disclose.

**Data Availability**: The data in my manuscript can be obtained from the first author upon reasonable request.